\documentclass[%
reprint,
amsmath,amssymb,
prl,
]{revtex4-1}

\usepackage{graphicx}% Include figure files
\usepackage{float}
\usepackage{dcolumn}% Align table columns on decimal point
\usepackage{bm}% bold math
\begin{document}

\preprint{APS/123-QED}

\title{Demonstration of cascaded modulator-chicane micro-bunching of a relativistic electron beam}

\author{N. Sudar, P. Musumeci, I. Gadjev, Y. Sakai, S. Fabbri}
\affiliation{Particle Beam Physics Laboratory, Department of Physics and Astronomy University of California Los Angeles\\Los Angeles, California 90095, USA
}%

\author{M. Polyanskiy, I. Pogorelsky, M. Fedurin, C. Swinson, K. Kusche, M. Babzien, M. Palmer }
\affiliation{
Accelerator Test Facility Brookhaven National Laboratory\\
Upton, New York 11973, USA% with \\
}%

\date{\today}% It is always \toda y, today,
% but any date may be explicitly specified

\begin{abstract}
We present results of an experiment showing the first successful demonstration of a cascaded micro-bunching scheme. Two modulator-chicane pre-bunchers arranged in series and a high power mid-IR laser seed are used to modulate a 52 MeV electron beam into a train of sharp microbunches phase-locked to the external drive laser. This configuration allows to increase the fraction of electrons trapped in a strongly tapered inverse free electron laser (IFEL) undulator to 96\%, with up to 78\% of the particles accelerated to the final design energy yielding a significant improvement compared to the classical single buncher scheme. These results represent a critical advance in laser-based longitudinal phase space manipulations and find application both in high gradient advanced acceleration as well as in high peak and average power coherent radiation sources.

\end{abstract}

\maketitle

Progress in the production of high brightness electron beams has provided the scientific community with a wide variety of tools for measuring phenomena at unprecedented spatial and temporal scales, making use of the short wavelength radiation generated by these beams or using the electrons as probe particles directly \cite{LCLS,UED2}. Enhancing the capabilities of these investigative tools has become an active area of research aimed at improving the peak and average brightness of the e-beam and the generated radiation, better controlling the spectral-temporal characteristics of the radiation, and decreasing the footprint of these devices using advanced accelerator techniques \cite{BBD}.  Many of these schemes demand precise control of the electron beam phase space at optical scales.

Laser-based manipulations of the electron beam longitudinal phase space can be achieved combining the sinusoidal energy modulation introduced when an electromagnetic wave and a relativistic electron beam exchange energy in an undulator magnet, with a dispersive element such as a magnetic chicane or a simple drift.  Modulator-chicane pre-bunching has been used to great effect, both for high efficiency generation of coherent radiation and in high gradient laser-driven acceleration \cite{HGHG, STELLA, Sears_PRSTAB, DurisNatComms, Nocibur}. The non linearity (sinusoidal dependence) of the energy modulation poses a limit on the quality of the bunching that can be achieved typically with only 60\% of the electrons contributing to the micro-bunch. Adding several of these elements in series with varying modulation strengths, laser wavelengths and dispersive strengths allows for complex tailoring of the energy and density distributions of the beam.  In this way it is possible to gain greater control of the electron beam harmonic content, peak current, current distribution, bunching factor, and output energy spread \cite{EEHG,ESASE,MLFEL,ATTO,OAB}.

For example, in seeded strongly tapered undulator interactions, particles gain or lose significant amount of energy to the radiation.  In order to maintain resonant interaction, one needs to modify the period and/or magnetic field amplitude along the undulator. When particles are injected near the initial resonant energy and phase, they are trapped in stable regions of longitudinal phase space called ponderomotive buckets, and follow phase space trajectories defined by the undulator tapering \cite{KMR}. By manipulating the initial energy-temporal profile of the electron beam, one can greatly increase the fraction of particles injected into these stable regions, maximizing the efficiency of this interaction\cite{dbt}.

In this letter, we present the results of an experiment successfully demonstrating one such longitudinal phase space manipulation whereby using two modulator chicane pre-bunchers in series we are able to produce a sequence of sharp spikes in the electron beam density profile, periodically spaced at the wavelength of a mid-IR seed laser.  Subsequently injecting these micro-bunches into the periodic stable ponderomotive potential of a strongly tapered undulator interaction in the accelerating configuration \cite{Palmer,CPZ,DurisNatComms,MoodyLLNL}, driven by the same seed laser pulse, we are able to trap and accelerate 96\%  of a 52 MeV electron beam, with 78\% of the particles reaching the final design energy of 82 MeV over the 54 cm undulator length within an RMS energy spread of 1\%.

\begin{figure*}[t]
\includegraphics[scale=0.6]{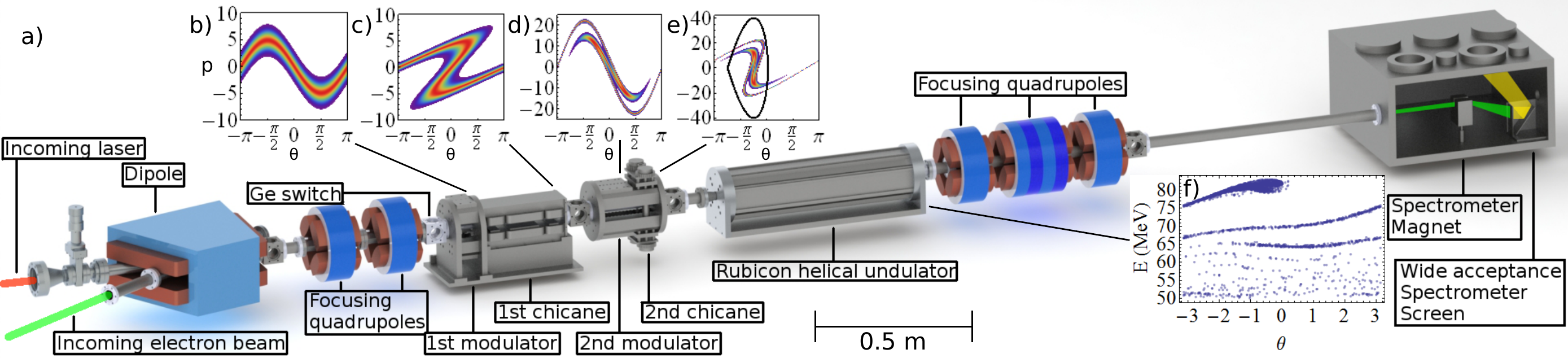}
\caption{\label{fig:wide}a) experiment beamline layout b)-e) electron beam phase space after first modulator, first chicane, second modulator, second chicane f) phase space after IFEL acceleration}
\label{fig:layout}
\end{figure*}

In order to understand quantitatively the benefits of the cascaded buncher configuration we can start from the equations describing the energy exchange between an electromagnetic wave and a relativistic electron beam copropagating in an undulator field (Eq. 1,2) \cite{Madey,Pellegrini}.
\begin{align}
\frac{d\gamma^2}{dz} &= -k K_l K JJ \sin(\theta)\\
\frac{d\theta}{dz} &= k_w - \frac{k(1+\frac{K^2}{2})}{2\gamma^2} = 0 \rightarrow \gamma_r^2 = \frac{k(1+\frac{K^2}{2})}{2k_w}
\label{energy}
\end{align}
\noindent where $k_w$ and $k$ are the undulator and laser wavenumbers, $K=\frac{eB_0}{k_w m_e c}$ and $K_l=\frac{eE_0}{km_ec^2}$  are the undulator and laser vector potentials, $JJ=J_0(\frac{K^2}{4+2K^2})-J_1(\frac{K^2}{4+2K^2})$, $\gamma$ and $\theta$ represent the particle Lorentz factor and phase respectively and $\gamma_r$ is defined as the resonant energy, with both expressions applying to a planar undulator geometry.

Over a short distance, i.e. one undulator period, we can ignore the phase evolution resulting in a sinusoidal energy modulation with a modulation amplitude of $\Delta \gamma \sim -k K_l K JJ z/2\gamma_r $.  Transformation of the scaled phase space variables, $p \equiv \frac{\gamma-\gamma_r}{\sigma_\gamma}$ and $\theta \equiv k z$, where $\sigma_\gamma$ is the electron beam initial energy spread, gives $p'=p+A \sin(\theta)$ and $\theta'=\theta$, where $A= \Delta \gamma/\sigma_\gamma$.

This energy modulation can be transformed into a density modulation using a dispersive element such as a 4-dipole magnetic chicane, transforming the phase space variables to, $p''=p'$ and $\theta''=\theta'+B[p+A\sin(\theta')]$, where $B \equiv \frac{R_{56}k\sigma_\gamma}{\gamma_r}$ with $R_{56}$ being the dispersive strength of the chicane.  Choosing $B \sim \frac{\pi}{2A}$ will produce a series of sharp spikes in the electron beam density distribution, rotating particles in the linear region of the energy modulation, between $-\frac{\pi}{2}<\theta<\frac{\pi}{2}$, to approximately the same phase.  If the induced energy modulation amplitude is less than the height of the ponderomotive potential, then a large fraction of the beam charge can be trapped by injecting these micro-bunches at the resonant phase and resonant energy of a tapered undulator.
\begin{figure}[b]
\includegraphics[scale=0.5]{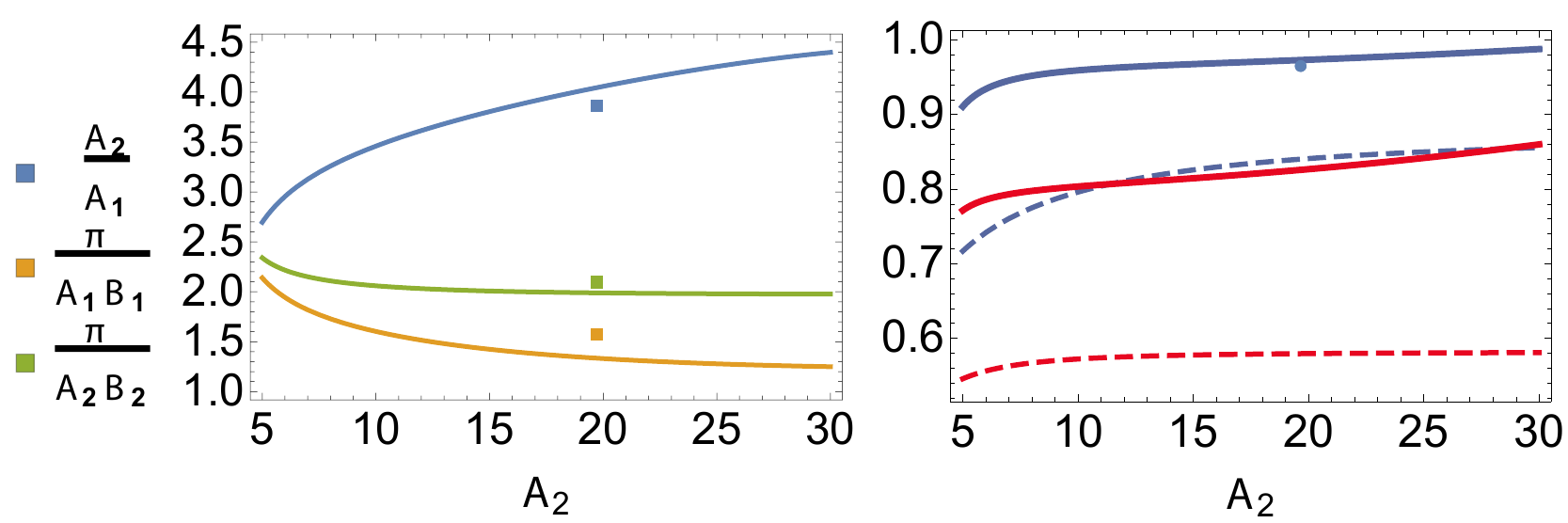}
\caption{(left) optimal double buncher parameter values maximizing trapping and experimental optimal values at $A_2=20$ (points). (right) Double buncher (blue, dashed) and single buncher (red, dashed)  fraction trapped and bunching factor. Point at $A_2=20$ shows measured trapping.}
\label{fig:parameters}
\end{figure}

Adding a modulator chicane module before the final pre-buncher serves to greatly increase the number of particles in the linear region of the final sinusoidal energy modulation, as shown in Figure \ref{fig:layout}d \cite{Hemsing}.  This scheme can be described defining the modulation amplitudes and scaled dispersive strengths, in order, as $A_1$, $B_1$, $A_2$ and $B_2$.  The first modulator imparts a small sinusoidal energy modulation with amplitude $A_1$, Fig. \ref{fig:layout}b.  The first chicane over-compresses this modulation, maximizing the number of particles between $-\frac{\pi}{2}<\theta<\frac{\pi}{2}$, with $B_1 \sim \frac{\pi}{A_1}$, Fig. \ref{fig:layout}c.  The second modulator imparts a sinusoidal energy modulation, now with a much larger fraction of particles in the linear region, Fig. \ref{fig:layout}d. The second chicane rotates this energy modulation into density modulation with $B_2 \sim \frac{\pi}{2A_2}$, Fig. \ref{fig:layout}e.

Fig. \ref{fig:parameters}a shows polynomial fits to the values of $A_1$, $B_1$, and $B_2$ which optimize this scheme for a given $A_2$, considering the laser energy to be a free parameter. The figure of merit for the manipulation depends on the final application. For radiation generation the bunching factor, $b = <e^{i \theta}>$ is often used. For advanced accelerators, one needs to maximize the trapping into a ponderomotive bucket. In this case, assuming the same laser beam drives the tapered undulator interaction with a resonant phase  $\theta_r=-\frac{\pi}{4}$, the energy acceptance of the application bucket is $A_b \propto \sqrt{K_l(\cos[\theta_r]+(\frac{\pi}{2}+\theta_r)\sin[\theta_r])}$.
The optimization of the relative strength of dispersion and modulation are reported in Fig. \ref{fig:parameters}a showing $B_1$ and $B_2$ converging to the qualitative estimates at large $A_2$.

These optimal values show an increase in the estimated trapping fraction from 80\% with a single pre-buncher to 98\% with the cascaded pre-bunching scheme and an increase in the bunching factor, $b$, from 0.6 to 0.8, Fig. \ref{fig:parameters}b.  The points in Fig. \ref{fig:parameters}a show parameters corresponding to our experimental setup. Considering initial trapping in the ponderomotive potential of the Rubicon IFEL, with $\theta_r=-\frac{\pi}{4}$ and experimental laser parameters giving $A_2 = 20$ and $A_b = 40$, yields an estimated trapping fraction of $\sim96\%$.

Figure 1a shows a schematic of the beamline at Brookhaven National Laboratory Accelerator Test Facility (ATF) with both modulator chicane pre-bunchers and the Rubicon helical undulator.  The experiment utilized a single high power pulse from a 10.3 $\mu m$ wavelength $CO_2$ laser with a pulse duration of several pico seconds to drive the interactions in both modulator chicane modules and the IFEL. The laser pulse is focused using a 4 m focal length NaCL lens to a 1.06 mm waist at the center of the undulator.  Including transport losses the laser power delivered to the IFEL fluctuates between 70-100GW.  Experimental parameters are listed in Table 1.

The electron beam is coaligned to the seed laser propagation axis after passing through a dipole and is then focused by a quadrupole doublet, maintaining a small electron beam cross section compared to the laser.  Picosecond scale timing between laser and electron beam arrival time is achieved first utilizing electron-beam controlled transmission of the mid-IR pulse in a semiconductor (Ge) slab \cite[]{Cesar:GeSwitch} and then optimized by maximizing the energy modulation on the electron spectrometer.  Both modulator chicane modules could be removed and inserted on the beamline without noticeable alignment errors, allowing for separate optimization of each component.  

The first modulator consists of a half period planar Halbach undulator with period 7 cm.  This is followed by an electromagnetic chicane consisting of 4 dipole electromagnets of length 3 ~cm separated by 3 cm drifts with a field of  2.25 mT/A over the range of 0-150 A, corresponding to $R_{56} =0-900 \mu$m, Figure 3b.  The second modulator consists of a single period planar Halbach undulator with period 5 cm.  This is followed by a variable gap permanent magnet chicane composed of 4 dipole magnets of length 12.5~mm whose gap can be adjusted from a minimum of 15.9~mm to a maximum of 22~mm and are interspaced by drifts of 12.5~mm, corresponding to $R_{56} =40-90 \mu$m, Figure 3a.  The Rubicon helical undulator is made up of two $N_w = 11$ period Halbach undulators, oriented perpendicularly and shifted in phase by $\pi$/2 with period increasing from 4.04 cm to 5.97 cm. The resonant energy of the undulator is tuned from 52 MeV at the entrance to a final energy of 82 MeV.

The use of a half period modulator, diffraction of the laser, and offset of the laser-electron beam timing, $\sim$0.75 ps, to compensate for the first chicane large delay, gives a modulation ratio, $\frac{A_2}{A_1}=3.9$. The laser pulse is circularly polarized to drive the helical undulator interaction. The use of planar modulators, again combined with diffraction effects, increases the modulation to bucket height ratio to $\frac{A_b}{A_2}=2$. The total slippage after both modulator-chicanes is $\sim$ 0.9 ps. The fine timing is adjusted to compensate for temporal slippage so that the electron beam enters the IFEL near the peak field of the laser.  This allows for additional control of the field amplitude at each component.
\begin{table}
\caption{\label{parameters} Parameters for the Rubicon experiment.}
\begin{ruledtabular}
\begin{tabular}{lcdr}
\textrm{Parameter}&
\textrm{Value}\\
\colrule
Initial electron beam energy & 52 MeV \\
Initial beam energy spread ($\frac{\Delta \gamma}{\gamma}$)& 0.0015\\
electron beam emittance ($\epsilon_{x,y}$)&2.5 mm-mrad\\
electron beam waist ($\sigma*_{x,y}$)&80$\mu$m\\
electron beam length ($\sigma_{z}$)&1 ps \\
electron beam charge & 80 pC \\
Laser wavelength & 10.3$\mu$m\\
Rayleigh range & 0.34 m \\
Laser waist & 1.06 mm \\
Waist position (undulator entrance @ z=0)&  0.16 m\\
Laser Power & 70- 100 GW\\
\end{tabular}
\end{ruledtabular}
\end{table}
\begin{figure}[t]
\includegraphics[scale=0.42]{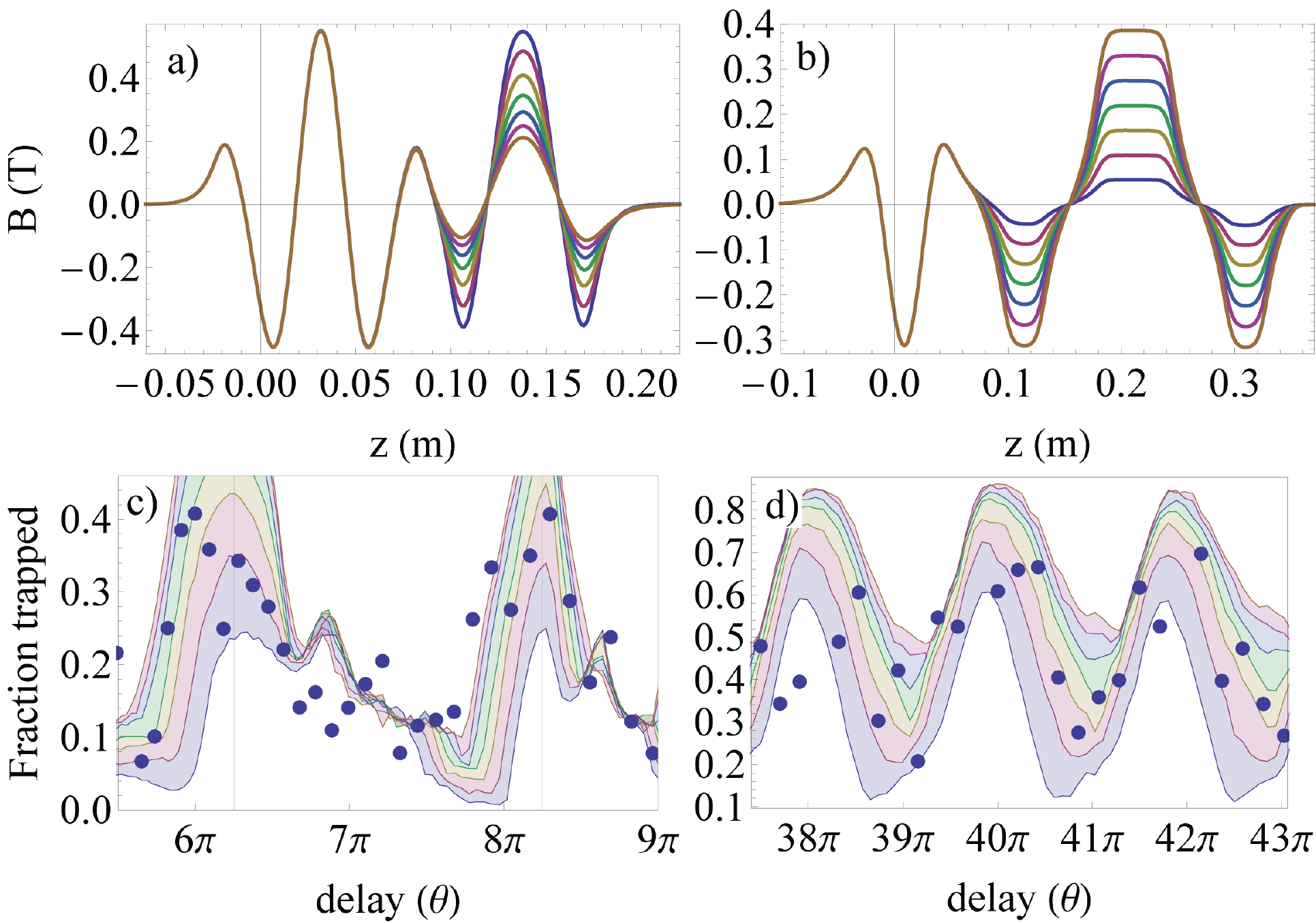}
\caption{a),b) Hall probe measurements of magnetic field profile of downstream pre-buncher varying chicane gap and upstream pre-buncher varying chicane current. c),d) Experimental data (dots) showing fraction accelerated $>77$MeV varying downstream pre-buncher chicane gap and upstream pre-buncher chicane current to control dispersion and injection phase compared with GPT simulations with laser power 70-100GW in steps of 8 GW.}
\label{pb_capture}
\end{figure}

The adjustable field of both chicanes allows for both tuning of the optimal $B_1$ and $B_2$ and precise control of the relative injection phase between the laser and the electron microbunches at both the second modulator and the undulator entrance.  In the experiment we first optimize $B_2$, inserting the downstream modulator chicane module, and scanning over the variable chicane gap, Figure 3c.  We observe two peaks in the fraction of particles trapped corresponding to phase slippages $S = 6\pi+\pi/4$ and $S = 8\pi + \pi/4$, corresponding to injection at the expected resonant phase, $-\pi/4$.

In order to refer to the discussion of Fig. \ref{fig:parameters}, we observe that $B_2 \sim\frac{2 \sigma_\gamma S}{\gamma_r}$ where the extra factor of 2 results from the relation between slippage and dispersion. The peak at $33\pi/4$ delay matches closely the analytical estimate for optimal $B_2$ with $\frac{\pi}{A_2 B_2}\sim2.1$ for $A_2=20$.  Setting the downstream chicane at the larger delay, we insert the upstream modulator chicane module and optimize $B_1$, varying the current in the chicane, Figure 3d.  We observe 2 peaks in the fraction of particles trapped corresponding to phase delays near $40\pi$ and $42 \pi$, corresponding to injection at 0 phase offset at the entrance of the second buncher, with a total dispersion again matching closely to the analytical estimate for optimal $B_1$ with $\frac{\pi}{A_1 B_1}\sim1.4$.

Simulating the interaction with General Particle Tracer (GPT) \cite{GPT} using electron beam and laser parameters measured experimentally and field maps from the 3D magnetostatic solver Radia \cite{Radia}, which agree well with Hall probe measurements of the undulator and both pre-bunchers, shows good agreement with experimental results.

\begin{figure}[t]
\includegraphics[scale=0.7]{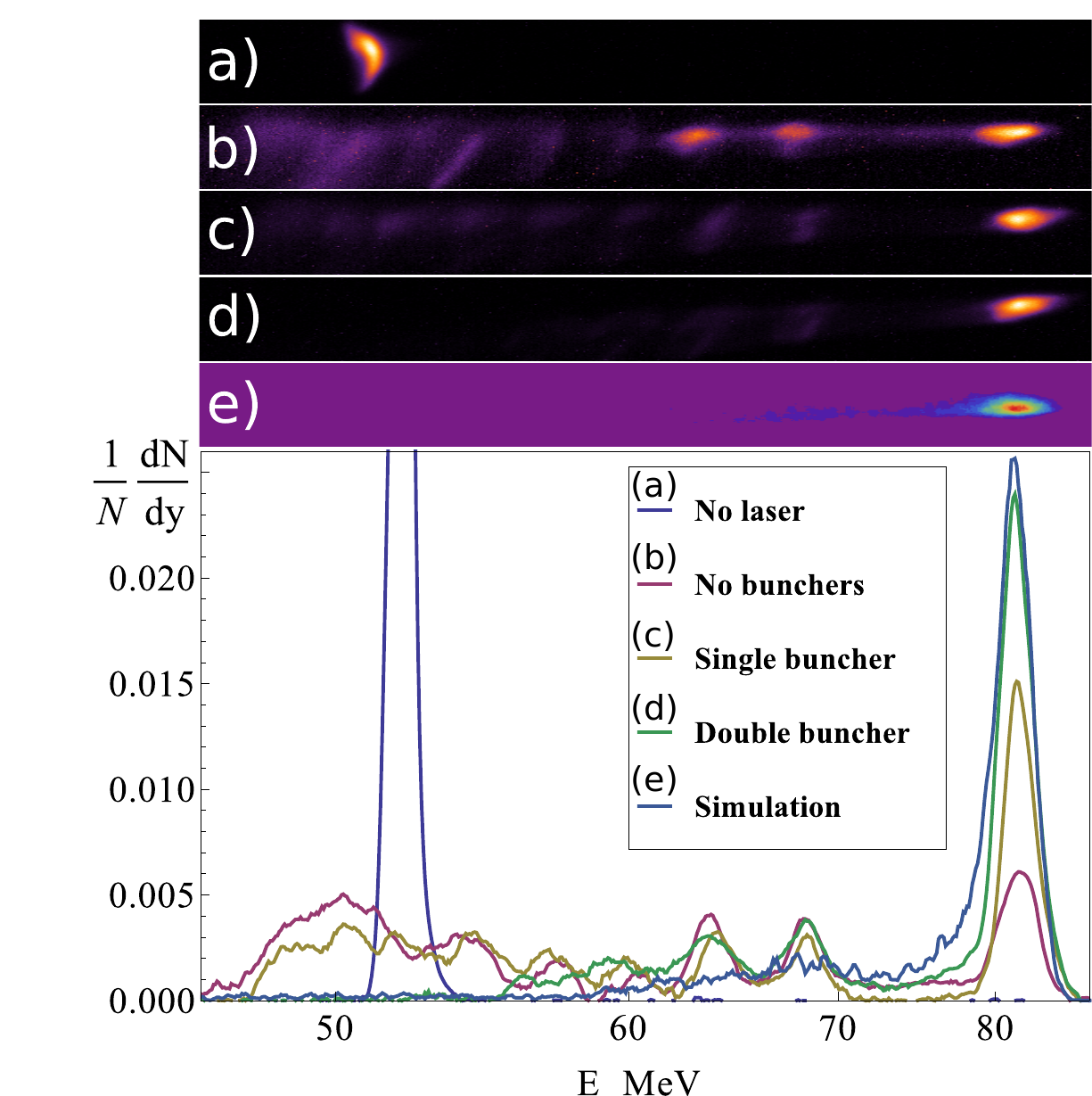}
\caption{a)-e) Raw spectrometer images with no laser seed, no bunchers, downstream buncher only, both bunchers and GPT simulation. (Bottom) Projections showing electron beam energy distribution $\frac{1}{N}\frac{dN}{dy}$ vs. $E$}
\label{spectra1}
\end{figure}

\begin{figure}[h!]
\includegraphics[scale=0.6]{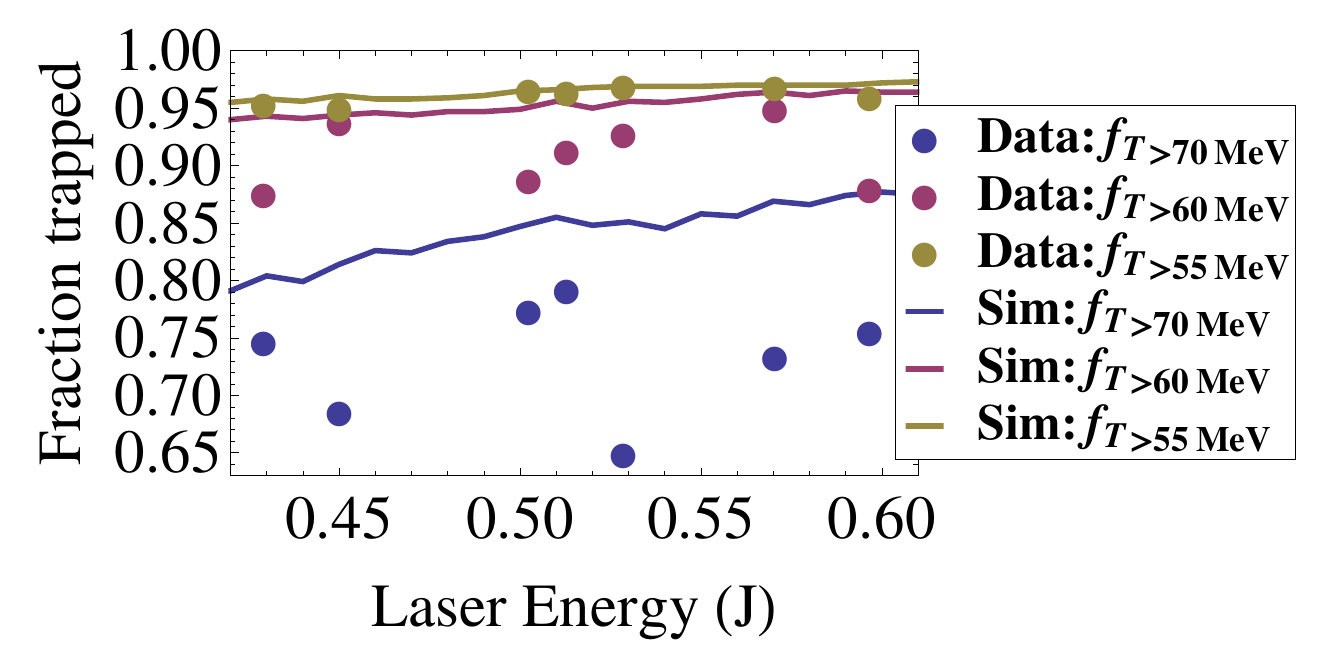}
\caption{Fraction accelerated $>55$MeV (Yellow), $>60$MeV (Red), $>70$MeV (Blue) comparing GPT simulations (lines) and experimental data (points) varying the laser power.}
\label{trap}
\end{figure}

In Fig. \ref{spectra1} we show energy spectrometer images taken with no laser, no bunchers installed, the downstream buncher installed, both bunchers installed and their relative distribution projections $\frac{1}{N_{tot}}\frac{dN}{dy}$ normalized so that the integral under the curves is 1.  The fraction of electrons accelerated past 77 MeV was $25\%$ for no bunchers, $40\%$ for single buncher, and $70\%$ for the double buncher, with all spectra taken from the same experimental run with a nominal 75 GW laser seed power.  Again we observe good agreement with GPT simulations,  Fig. \ref{spectra1}e.  The energy spread of the accelerated beam is $\sim1\%$ set by the amplitude of the ponderomotive bucket at the exit of the undulator. 

Figure \ref{trap} shows, for a series of laser shots of varying energy, $\sim96\%$ of the particles are accelerated past the initial energy, agreeing well with GPT simulations and analytical estimates.  The decreased trapping fraction, $f_T$, for particles accelerated above 60 MeV and 70 MeV can be attributed to non ideal electron beam and laser focusing in the experiment.  Errors in the electron beam trajectory in the undulator contribute to increased detrapping throughout the interaction, manifesting in increasing discrepancy between simulation and data at higher energies. The demonstration of nearly complete initial trapping of the electron beam provides a clear validation for the cascaded bunching scheme.

In conclusion, the cascaded pre-buncher Rubicon experiment demonstrated initial trapping of $96\%$ of a 52 MeV electron beam with up to 80\% of the electron beam reaching the final energy of 82 MeV, decreasing the number of initially detrapped particles by an order of magnitude compared to the single buncher case. These results agree well with both simulations and analytical estimates.  This experiment took advantage of the favorable parameters of the $CO_2$ laser, characterized by long pulse lengths and energies on the order of 1 J, allowing use of a single laser pulse to drive the entire interaction without great effect from the electron beam slippage from the modulator chicane elements.  The long wavelength of the $CO_2$ laser also allows for increased stability and phase space acceptance. Scaling this scheme to higher energy electron beams and shorter wavelength laser seeds, which typically exhibit much shorter pulse lengths, may require particular care in controlling the relative slippage between electron beam and radiation and phase-locking between the different stages. The use of separate laser pulses to drive each modulator chicane might offer a greater range of tunability between the modulation and dispersion strengths allowing for a large variety of schemes to be investigated. Using a larger ($>2$) number of modulator-chicanes in series can lead to further enhancement of the micro-bunching, theoretically producing a bunching factor $>$0.9.

Successful demonstration of this scheme not only increases the performance of IFEL accelerators and their applications, but also encourages exploration into other areas where cascaded pre-bunching could prove useful.  This includes schemes where cascaded pre-bunching could be used to increase the efficiency of a strongly tapered FEL, to greatly increase the electron beams peak current for enhanced self amplified spontaneous emission in an FEL, or to excite resonances in dielectric structures \cite{GA}.  Recent demonstration of fresh bunch self seeding, for example, offers a unique scenario where a high power monochromatic x-ray seed and an electron beam with small slice energy spread are available \cite{CE}, providing a situation where cascaded pre-bunching could be employed to great effect.

\begin{acknowledgments}
This work was partially supported by DOE grant No. DE-SC0009914, U.S. DHS DNDO under Contract No. 2014-DN-077-ARI084-01 and the DOE SCGSR program. The authors would also like to thank Gerard Andonian for useful comments, Phuc Hoang and Ryan Roussel.
\end{acknowledgments}

\end{document}